\documentclass[12pt,preprint]{aastex}

\newcommand{\source}{{\it S5 0716+714 }}

\shorttitle{Brightest flaring event in BL Lac Object S5 0716+714}
\shortauthors{Chandra et al.}

\begin{document}

\title{Multi-wavelength study of flaring activity in BL Lac object S5 0716+714 during 2015 outburst}

\author{Chandra, Sunil\altaffilmark{1}, Zhang, Haocheng\altaffilmark{2,5}, Kushwaha, Pankaj\altaffilmark{1}, Singh, K. P.\altaffilmark{1}, Bottcher, M.\altaffilmark{3}, Kaur, Navpreet\altaffilmark{4} \& Baliyan, K. S.\altaffilmark{4} }

\affil{Department of Astronomy \& Astrophysics, Tata Institute of Fundamental Research,
    Mumbai 400005, India}

\altaffiltext{1}{Tata Institute of Fundamental Research, Mumbai 400005, India}
\altaffiltext{2}{Astrophysical Institute, Department of Physics and Astronomy, Ohio University, Athens, OH 45701, USA}
\altaffiltext{3}{Centre for Space Research, North-West University, Potchefstroom 2531, South Africa}
\altaffiltext{4}{Physical Research Laboratory, Ahmedabad 380009, India}
\altaffiltext{5}{Theoretical Division, Los Alamos National Laboratory, Los Alamos, NM 87545, USA}
\begin{abstract}
We present a detailed investigation of the flaring activity observed from a BL Lac object, \source, during its brightest ever optical state in the second half of January 2015. Observed almost simultaneously in the optical, X-rays and $\gamma$-rays, a significant change in the degree of optical polarization (PD) and a swing in the position angle (PA) of polarization were recorded. A detection in the TeV (VHE) was also reported by the MAGIC consortium during this flaring episode. Two prominent sub-flares, peaking about 5-days apart,  were seen in almost all the energy bands. The multi-wavelength light-curves, spectral energy distribution (SED) and polarization are modeled using the time-dependent code developed by \citet{Haocheng2014}. This model assumes a straight jet threaded by large scale helical magnetic fields taking into account the light travel time effects, incorporating synchrotron flux and polarization in 3D geometry. The rapid variation in PD and rotation in PA are most likely due to re-connections happening in the emission region in the jet, as suggested by the change in the ratio of toroidal to poloidal components of magnetic field during quiescent and flaring states.        
\end{abstract}


\section{Introduction}

 Blazars are an extreme subclass of active galactic nuclei (AGN) known to posses an extremely collimated relativistic jet, perpendicular to the plane of the accretion disk and seen at very small angles (\textless $15^\circ$) to our line of sight (LOS) \citep{UrryPadovani1995,UrryPadovani2000}. Such close alignment of the jet to the LOS leads to the relativistic boosting of the jet emission, which dominates the blazar emission. The spectral energy distribution (SED) of a blazar typically exhibits bi-modality with two broad bumps: one peaks between sub-mm to UV/X-rays, and the other peaks somewhere at the MeV-GeV energies. The low energy part of SED is well established to be due to synchrotron process in the relativistic jet \citep{UrryMushotzky1982}, with other non-jet components like disk, torus, BLR etc. also being significant contributors in several sources albeit in narrow energy range \citep{Ghisellini2009, Nalewajko2014, Kushwaha2014}. The high energy bump, on the other hand, is poorly understood and according to one approach, this component is thought to arise from inverse Compton scattering of low energy seed photons by the highly energetic leptons ($e^+/e^-$) in the jet. If the synchrotron photons, originating in the same population of high energy leptons are upscattered to high energies, the process is termed as synchrotron self Compton (SSC) \citep{BloomMarscher1996,Sokolov2004,Ghisellini1985}. In the cases where the seed photons are external to the jet, namely, disk, torus, BLR and/or sometimes even CMBR, then the process is known as external Comptonization \citep{Dermer1992,Sikora1994,Sikora2009,Agudo2011}. 
 
  The emission from blazars shows enormous amount of variability at almost all the frequencies. Since the central engine is not resolvable by any existing facility, the variability provides a useful tool to diagnose the physical mechanisms responsible for the emission, thanks to the availability of quasi-simultaneous data at various energies from space-based (e.g. {\it Fermi, Swift etc. }) and many ground based observatories. Optical polarization observations can sufficiently constrain many of the jet properties e.g., the strength and nature of the magnetic field, the geometry of the jet, and the physical processes etc., provided these observations are properly supplemented by the variability information at other energies. The high energy $\gamma$-ray flares are mostly followed by the activities at lower energies with a few exceptions of orphan flares. The observability of the emission at a particular energy from a blazar jet may be dependent on the opacity of the emission region at that wavelength. Therefore, in this scenario, the location of the dissipation region becomes an important factor behind the nature of variability seen across the electromagnetic spectrum. In other words, a multi-wavelength variability study of blazars can also provide us information about the location of the emission region in the jet \citep{Marscher2010,Orienti2013}. 
  
  In the literature, some cases of rotation or swing in PA have been reported during high energy flares in blazars \citep{Abdo2010Nature,Marscher2014,Haocheng2014}. Being dependent only on the orientation of shock and magnetic field threading it, PA provides an unique tool to understand the acceleration mechanisms and the behavior of the shocked plasma. Several models have been proposed to understand the PA swings, however, currently only two of them are capable of reproducing the multi-wavelength light-curves, time-dependent SEDs, and multi-frequency polarization, simultaneously. The first model, known as the Turbulent, Extreme Multi-zone Model (TEMZ), suggested by \citet{Marscher2014}, assumes that the emission region comprises of turbulent chaotic magnetic field with a large number of small regions of ordered small scale magnetic fields called `cells'. A proper accounting of time dependent contributions of individual cells to the total emission is done to simulate the observables. The second approach termed as the Helical Magnetic Field Model (HMFM), suggested by \citet{Haocheng2014} assumes a large scale ordered helical magnetic field with a more rigorous and proper accounting of the Light Travel Time Effects (LTTEs) for individual zones to recreate the SEDs, light-curves and polarization. 

The blazar \source, at a red-shift, z of 0.31 \citep{Nilsson2008} is one of the brightest BL Lac objects that is highly variable from radio to $\gamma$-rays with a very high duty cycle \citep{WagnerWitzel1996}. In the SED based classification scheme, \source is sub-classified as an intermediate energy peaked BL Lac object (IBL) \citep{Giommi1999}, and is also confirmed based on the concave shape of its X-ray (0.1-10.0 keV) spectrum \citep{Foschini2006, Ferrero2006}. The concave shape of the spectrum in X-rays is indicative of the presence of a tail from the synchrotron emission (falling tail) and a flatter part from the inverse Compton (IC) spectrum (rising tail). This object has been detected in the MeV energy range several times at different flux levels by the Energetic Gamma Ray Experiment Telescope (EGRET) detector on board the Compton Gamma-ray Observatory ({\it CGRO}) \citep{Hartman1999,Nandikotkur2007}. In 2008, Astro‐Rivelatore Gamma a Immagini Leggero ({\it AGILE}) reported the detection of variable $\gamma$-ray flux with a peak flux density above the maximum reported by EGRET \citep{Chen2008}. \source is also in the Fermi-LAT bright source list \citep{Abdo2009}. {
 MAGIC first detected this source in VHE $\gamma$-rays during 2007 November ($F_{>0.4TeV} \sim 0.8 \times 10^{-11} \rm ~erg~cm^{-2}~ s^{-1}$) and later with much larger flux ($F_{>0.4TeV} \sim 7.5 \times 10^{-11} \rm ~erg~cm^{-2}~ s^{-1}$) at 5.8 $\sigma$, in 2008, April. Similar trend was also seen in optical \citep{Anderhub2009} while source was in historically high state in X-rays \citep{Giommi2008} in 2008, April. A concurrent rapid rotation of the PA was also observed just after the maximum in optical flux had reached  \citep{Larionov2008}. This seems to support the indication seen in the previous MAGIC observations for other BL Lac objects \citep{Albert2006, Albert2007a}, that there is a connection between the optical high states and the VHE $\gamma$ -ray high states.

  The recent optical monitoring of \source shows a consistent rising trend in the R band flux over the past few months. The older observations reveal that the source had gone to a very faint state (14.9 mag in R-Band) around 2013, December (MJD 56650), very close to the faintest state of this source reported till date (private communications). Very recently, \citet{Carrasco6902Atel} had reported a high state in IR on MJD 57033.3 (2015, January 11) which was around 2.5 magnitude brighter than the previously observed flux on MJD 57021 (2014, December 29). Following these observations, there were several telegrams reporting a further brightening of \source in different optical and IR bands \citep{Arkharov6942Atel, Bachev6944Atel, Bachev6957Atel, Spiridonova6953Atel, Chandra6962Atel}. Soon after these reports of optical/IR flares,  \citet{Mirzoyan6999Atel} reported a variable VHE detection above 150 GeV. This VHE detection, quasi-simultaneous with a high flux state in the optical, seems to be similar to that seen in the 2008 event. In the present work, we have investigated this event in the framework of time-dependent modeling of the observables, namely, light-curves, SEDs and multi-frequency polarization, to understand the role of the magnetic field in the blazar jet. Our study is mainly focused on the simulation of the part of the outburst where simultaneous data at all energies (optical to $\gamma$-rays) are available. This paper is organized as follows: \S 2 provides the details of data resources and the analysis methodology adopted. We present the multi-wavelength light-curves and SEDs along with the modeling of these in \S 3, followed by the discussion of the results obtained, in \S 4. Finally, in  \S 5 we present a summary of our work.

\section{Observations \& Data Analysis}

We used data from Large Area Telescope (LAT), on-board {\it Fermi} spacecraft \citep{Atwood2009}, for $\gamma$-ray counterparts of the optical flaring event. The {\it Swift}/XRT \citep{Burrows2005} and {\it Swift}/UVOT \citep{Roming2005} data are analyzed for X-rays and UV light-curves and SEDs. The optical monitoring data from 1.2 m \& 0.5 m telescopes at Mt. Abu Infra-Red Observatory (MIRO) are also analyzed. A few other observations, reported in ``The Astronomers Telegrams", are also used in the present study. The corresponding values are corrected for galactic extinction and reddening. The publicly available spectropolarimetric observations from Steward Observatory, Arizona, are also used. In the following, we summarize the details of the observations and data analysis techniques used for various data sets.

\subsection{Fermi-LAT}  

LAT normally works in all sky scanning mode, covering the whole sky every 3 hours. The scanning mode along with its large field of view (FOV) provide almost 30 min of monitoring of each source during the course of each scan. The broad energy coverage of LAT ($\Delta E \approx$ 0.02-300 GeV) makes this facility ideal for studying high energy astronomical events. 

   Two months of Fermi-LAT {\it P7REP} data from 2015, January 01 to 2015, March 01 were analyzed using the Fermi ScienceTools {\it v9r33p0}. In the analysis, only SOURCE class events with energies between 100 MeV - 200 GeV, defined under the Instrument Response Function (IRF) {\it P7REP{\_}SOURCE{\_}V15}, have been used from a $15^\circ$ region of interest (ROI) centered at the location of S5 0716+714 (R.A.=$110.473^\circ$, DEC.=$+71.343^\circ$). We have not included the low energy and high energy ends of LAT data to avoid the possible contamination with artificial counts caused by the poor response of the detector in this range. All the sources from 2FGL catalog \citep{Nolan2012} within the ROI, plus an annular radius of $10^\circ$ around it, were modeled along with the standard diffuse templates of Galactic ({\it gll{\_}iem{\_}v05{\_}rev1.fit}) and isotropic background ({\it iso{\_}source{\_}v05{\_}rev1.txt}). For generating the XML model file, we have made use of the contributory python package, make2FGLxml.py\footnote{\url{http://fermi.gsfc.nasa.gov/ssc/data/analysis/user/}}. The criterion for using a proper model for generating light-curves and SEDs is similar to the one adopted by \citet{Chandra2014} and \citet{Kushwaha2014}. At first, all the sources within ROI+$10^\circ$ were considered for unbinned likelihood analysis. The point sources with TS values less than 0, obtained from likelihood analysis, were removed from our input model. The likelihood analysis procedure was repeated until it converged without any source with TS $< 0$. The daily fluxes for the lightcurve were then extracted using the best model parameters where both spectral indices and normalization were kept free. However, for generating the SEDs, spectral indices in the model were kept fixed at -1.6 $\pm$ 0.01 as obtained from the likelihood analysis of the complete data set.    

\subsection{Swift Data Analysis}
The 0.3-10.0 keV X-ray and UV/optical archival data from XRT and UVOT instruments, on-board the {\it Swift} X-ray space-borne observatory \citep{Gehrels2005}, were analyzed for the present work. The standard procedures prescribed by the instrument teams were followed step by step to generate the science products. The recently updated version of the calibration database (CALDB) along with {\it heasoft v6.16} were used for our analysis. A number of pointings made by {\it Swift} during 2015, January 19-31 were analyzed. The following is a quick summary of the analysis procedures used for XRT and UVOT data. 
\subsection{XRT Data}
 The level 2 cleaned event files were generated using the standard {\it xrtpipeline} tool with the default parameter settings and by  following the prescriptions of the instrument team. The source and background lightcurve and spectra were generated with appropriate region and grade filtering using the {\it xselect} tool. In this case, the source spectrum was extracted for a circular region of $47''$ radius around the source location while four source-free regions in the neighborhood of the target, each with a $100''$ radius, served for the background spectrum. The count rate in these observations exceeded the recommended pile-up free count rate for PC mode (0.5 counts/sec). Therefore, all  the event files were investigated for pile-up and a proper procedure for pile-up correction as suggested by other researchers\footnote{http://www.swift.ac.uk/analysis/xrt/pileup.php} were used wherever needed. The ancillary response matrix was generated using the task {\it xrtmkarf} followed by the {\it xrtcentroid} task. The response matrix file provided with the CALDB distribution was used for further analysis. 

  Spectral fitting was done in the energy band between 0.3 to 10.0 keV using the XSPEC (version 12.8.2) package distributed with {\it heasoft} package. A simple power law along with the Galactic absorption gives the best fit for almost all the observations of interest. The model parameter, $N_H$ i.e., the  interstellar column density, was kept fixed at a value of 3.06 $\times$ $10^{
20}$ $cm^{-2}$ \citep{Kalberla2005}. This value was estimated using the web-based tool developed by a group at the University of Bonn\footnote{https://www.astro.uni-bonn.de/hisurvey/profile/index.php}. The normalization and spectral index of the power law were the free parameters for the spectral fitting. Table 1 summarizes the values of various parameters obtained from the spectral fitting for different observation IDs. The 0.3-10.0 keV fluxes thus obtained were used for constructing the light-curves. 

 The Galactic absorption corrected X-ray energy spectrum was constructed using the procedure adopted in \citet{Chandra2014} with that instead of using the default binning, all spectra files were binned according to a fixed input file (describing the details of channel binning) using the {\it grppha} tool. The multiple X-ray spectra, if any, for a particular SED segment, were merged keeping in mind the possible spectral variations.       
 
\subsection{UVOT Data}   
UVOT snapshots with all the six available filters, V (5468 $\AA$), B (4392 \AA), U (3465 \AA), UVW1 (2600 \AA), UVM2 (2246 \AA), and UVW2 (1928 \AA) for all the ObsIDs were integrated with the {\it uvotimsum} task and analyzed using the {\it uvotsource} task, with a source region of $5''$, while the background was extracted from an annular region centered on the source location with external and internal radii of $40''$ and $7''$, respectively. The fluxes thus obtained, were corrected for galactic extinction using a tool developed for R platform. This interactive tool adopts the model described in \citet{Cardelli1989}. This tool needs E(B-V) value as input, which was estimated using the web-based calculator by NASA/IPAC Infrared Science Archive\footnote{http://irsa.ipac.caltech.edu/applications/DUST/} \citep{Schlafly2011}. The value of this parameter in the direction of S5 0716+714 is 0.026 $\pm$ 0.001.
 The extinction corrected fluxes obtained from the aforementioned procedures were then used for extracting the light-curves and SEDs.
\subsection{Optical Photometry \& Polarization Data}
 The photometric monitoring of S5 0716+714 was performed using iKon ANDOR CCD Camera (2048 $\times$ 2048) as a backend instrument at the f/13 Cassegrain focus of the 1.2m optical telescope of MIRO. The CCD in this camera is cooled to $-80^\circ$ with thermoelectric (TE) cooling to keep the dark current very low. The camera is attached to a coupling unit consisting of a filter wheel with 12 slots, equipped with 10 optical filters (Standard Johnson/Cousins UBVRI broadband filters + u,g,r Sloan filters + two narrow band filters) just on top of the camera. One of the two additional slots is kept blocked for grabbing bias/dark frames, whereas the other is kept open for white light monitoring of extremely faint sources. The bias and sky flats were taken on daily basis for performing the pre-photometric image processing.  
  The source was also monitored using the Automated Telescope for Variability Studies (ATVS) at MIRO mounted with TE cooled ($\sim -80^\circ$) ANDOR iXon EMCCD camera (1024 $\times$ 1024) as the backend instrument. This camera is coupled with a filter wheel with the Standard Johnson/Cousins UBVRI filters. The twilight sky flats and bias frames were also captured on daily basis. The observing strategy for both the facilities was to capture a few images in all the bands (UBVRI) and then monitor for a longer time in one filter, say R band. 
   The apparent magnitudes of the source and the known field stars (photometric standards) for individual exposures were derived using the standard aperture photometry technique preceded by pre-processing of images \citep{Chandra2011,Chandra2014}. The magnitude was then converted into flux using simple conversion factors and zero point flux (ZPF) for different bands \citep{Bessell1979PASP}. Apart from photometric monitoring at MIRO, the observed fluxes from various Astronomers Telegrams were also used for comparison and completeness. The R band fluxes obtained from the Steward observatory database are also used followed by extinction correction.    
   
   Optical polarimetric data used in this study were taken from the observations performed as a part of the Fermi Support observing Program at Steward Observatory, Arizona, USA \citep{Smith2009}. The PD and PA values provided at Steward observatory data base are already calibrated and hence can be readily used. We have corrected the PA values for $180^\circ$ angle ambiguity, in order to make the rotation clearly visible (Fig. \ref{fig1}(f)). 
%
\section{Results and Interpretation}
The multi-wavelength light-curves derived using observational data from various resources, mentioned in \S2 are shown in Fig. \ref{fig1}. The SED for the duration of MJD 57045.5-57047.5 was also generated using data from the aforementioned facilities. In the following, we discuss the results and their interpretations.

\subsection{Multi-wavelength light-curves}

  The Figs. 1(a) \& 1(b) show the $\gamma$-ray and X-ray fluxes observed by {\it Fermi} \& {\it Swift}, respectively. The Fig. 1(c) shows the UV/optical light-curves derived from {\it UVOT} observations while Fig. 1(d) shows R band magnitudes from various resources. The Figs. 1(e) \& 1(f) are the PD and PA observations from the Steward observatory. The figures show significant variability in all the energy bands ($\gamma$-rays to UV/optical) as well as in the optical polarization. As evident from Fig.  1(a), $\gamma$-ray show a clear trend of multiple ups and downs during the span of MJD 57034 to 57055. A consistent rise in flux is seen between MJD 57034.5 to 57039.5, reaching to a flux level of 0.75 $\times$ $10^{-6}$ ph $cm^{-2}$ $s^{-1}$, which is about 4 times the average $\gamma$-ray flux of this source. A rapid variation in flux is seen during MJD 57040.5 to 57042.6 reaching to a flux level of 0.83 $\times$ $10^{-6}$ ph $cm^{-2}$ $s^{-1}$ on MJD 57041.5, i.e., within one day, the flux increased and then decreased to 0.35 $\times$ $10^{-6}$ ph $cm^{-2}$ $s^{-1}$ the next day. The flux again started rising slowly after MJD 57042.5 and reached upto 1.1 $\times$ $10^{-6}$ ph $cm^{-2}$ $s^{-1}$, the highest $\gamma$-ray flux level reported for this source. However, a fast decrease in flux soon after this flare is also seen. We, therefore, notice two major $\gamma$-ray sub-flares associated with the January 2015 major flare in \source.    
  The {\it Swift} X-ray light-curve in 0.3-10.0 keV band (Fig. 1(b)) exhibits a consistently rising trend with a peak flux of $2.7 \pm 0.3$ $\times$ $10^{-11}$ erg $cm^{-2}$ $s^{-1}$, on MJD 57047.2, which is comparable to the peak X-ray flux seen during 2007 October-November flare of the \source \citep{Giommi2008}. The poor coverage of the XRT observations restricts us from making conclusions about multiple sub-flares in X-rays. However, the ups and downs in the individual data points provide a glimpse of small scale variations. 
  
  The R band optical light-curve (Fig 1(d)), shows two distinct, well separated sub-flares. Initially, the R band flux slowly rises and reaches to 11.6 mag from 12.41 mag during MJD 57035 - 57041.3 (rate $\sim$ 0.13 mag/day). The peak flux here corresponds to brightest ever state of the source reported so far \citep{Chandra2011}. Just after this peak, the flux decreases to 12.22 magnitude between MJD 57040.5 to 57044.4 (0.14 mag/days). Here, we have ignored the variations in subsequent nights. Specifically, the source had undergone very fast decrease in flux ($\Delta M$ $\sim$ 0.5 mag) during MJD 57040 to 57041.8 and then again went to 11.9 mag on MJD 57042.4 before falling to 12.2 mag on MJD 57044.2. Soon after, the R band flux again starts rising at a slightly faster rate (0.33 mag/day). The peak flux corresponding to the second bump is almost equal to that of the first one. However, the flux value of the second bump remains constant for quite some time ($\sim$ 2 days). After this, the flux gradually decreases to 12.68 mag within MJD 57046 to MJD 57054.5 (rate $\sim$ 0.1 mag/day). The later monitoring suggests even fainter state of this source. Note that the plotted flux values represent fluxes averaged over few hours of monitoring. 
   The fluxes in the other optical bands, namely, V, B and I, also show the similar behavior, but are omitted in the light-curves for the sake of clarity of Fig. \ref{fig1}. Fig. \ref{colvar} depicts the correlation between fluxes in different optical bands and the flux-dependence of the B-V color. The fluxes in various optical bands are differently correlated, which is clearly seen from the slopes of the best linear fits to these curves (Fig. \ref{colvar}). The second block of Fig. \ref{colvar} represents the standard {\lq bluer when brighter (BWB)\rq} trend seen during a typical flare in blazars. The {\it Swift UVOT} observations (Fig. 1(c)) also indicate the existence of two humps in the light-curve. However, the poor coverage by {\it Swift} pointing provides an incomplete picture of the flux variations. 
      
  The variations in the optical PD and PA are presented in the Figs. 1(e) \& 1(f), respectively. The PD shows various episodes of rapid variation with the trends completely unrelated to the total flux variations. More specifically, at the very beginning of our available dataset i.e., around MJD 57042, when \source had already passed through the first optical sub-flare, the PD was very high ($10.67 \pm 0.02$ \%), which then decreased to $4.0 \pm 0.02$ \% by MJD 57045.3. The next two observations reveal a 5.4 \% change in PD between two epochs in the same night differing by 2.4 hours (MJD 57045.3-57045.4). The next episode of PD is even more dramatic and nicely covered as the PD decreases by 4.68 \% within $\sim$ 7.2 hours (8.88 - 4.20 \% between MJD 57046.2 to MJD 57046.5). During the next three segments, namely, MJD 57047.2-57047.5, MJD 57048.1-57048.2 and MJD 57050.3-57051.5, independent rising trends of 4.4 \% (6.38-10.79 \%), 2.96 \% (8.0-10.96 \%) and 8.01 \% (0.45-8.46 \%), respectively, are seen. The typical error in these observations of PD is 0.02 \%. A straight line was fitted to the individual segments of PD variations using least square fitting algorithm, which clearly indicates very different slopes for all of them (L1 to L5 in left panel, Fig. \ref{pollinfit}). It is very difficult to associate rapid fluctuations in PD to the double-humped shaped variation in the total flux light-curve. It could perhaps be due to two emission components contributing to the total flux;  one nearly unpolarized and other polarized; both  varying with time. The observed PA profile also supports the same argument,​ as discussed below. Initially, during the first segment of the PD variations (MJD 57042.3-57045.3), the PA is mildly variable around $100^{\circ}$ with a slight change from $78.6^{\circ}$ to $100.2^{\circ}$. This implies the decay of a previous polarized component and the emergence of a new one; both with similar PA. Afterwards, the new polarized component starts to evolve. During MJD 57045 to 57047.5, the PA consistently increases by $\sim$ $164^{\circ}$, from $93.0^{\circ}$ to $256.9^{\circ}$. The later observations (MJD 57048.1 to 57051.5) suggest a consistent decrease in PA from $256.9 ^{\circ}$ to $96^{\circ}$. The rising and falling parts of the PA variations were fitted with an exponential function [$f(x) \sim A e^{-\alpha (x-x0)}$], to get an estimate of the temporal profile (right panel, Fig. \ref{pollinfit}). The index of the exponential for the rising part is $-0.45 \pm 0.01$ $day^{-1}$ whereas that for the falling part is $0.29 \pm 0.006$ $day^{-1}$. This shows that the second PA swing is slower than the first one (right panel, Fig. \ref{pollinfit}). These make the case for two rotations, of about $180^\circ$ each, during the course of the 2nd optical sub-flare, indicating the sub-flare to be strongly related to a significant change in the magnetic field. 

  \citet{Mirzoyan6999Atel} reported MAGIC detection of the variable VHE $\gamma$-ray flux in the range of 4 $\times$ $10^{-11}$ to 7 $\times$ $10^{-11}$ ph $cm^{-2}$ $s^{-1}$ above 150 GeV, during MJD 57044-57048, which is simultaneous with the X-ray and optical high states of the \source during the second sub-flare (Fig. \ref{fig1}). It appears that this variable VHE detection is  correlated with the activity in the optical and the X-rays, similar to the 2008 detection of VHE $\gamma$-ray emission of \source \citep{Anderhub2009}. In the following, we have described the modeling of the observables derived for the duration of MJD 57045.5-57047.5.

\subsection{Modeling of Light-curves, Optical Polarization \& SED}

  In general, there are two mechanisms, namely, shocks and magnetic reconnection, that may result in a flaring activity in a blazar jet. We have used the HMFM model proposed by \citep{Haocheng2014} to fit the first PA rotation, from MJD 57045.5 to 57047.5 (the rising part). This model assumes an axis-symmetric cylindrical geometry for the emission region, which is further evenly subdivided into zones in radial and longitudinal directions \citep[see Fig. 1,][]{Haocheng2014}. The relativistic plasma in the emission region, pervaded by a helical magnetic field with a possible additional turbulent component, moves on a straight trajectory along the jet and encounters a flat stationary disturbance. The disturbance will temporarily modify the physical conditions (magnetic field, particle distribution etc.) at its location inside the emission region. The initial state in the emission region is regained once it moves out of the shocked region. The non-thermal particle evolution, radiation and polarization signatures of our model are realized by the Monte-Carlo/Fokker-Planck (MCFP) radiation transfer code developed by \citet{Chen2012} and the 3D polarization dependent radiation transfer code developed by \citet{Haocheng2014}. In this model, the particle distribution in individual zones are evolved using a locally isotropic Fokker-Planck equation, and the polarization dependent emission from each zone is properly traced to account for all the LTTEs. Even if the disturbance in our model is flat, the flaring region, which consists of the zones affected by the disturbance whose emissions arrive to the observer at the same time, will be distorted into an elliptical shape. Illustrations of the model setup and the LTTEs can be found in Fig. \ref{modelsetup} and also in \citet{Zhang2015}.

Table 2 lists some key parameters from our modeling, estimated in the co-moving frame of the emission region. Due to the relativistic aberration, even though we are observing these objects at very close alignment to the relativistic jet in the observer's frame (typically, $\theta^{\ast}_{\rm obs} \sim 1/\Gamma$, where $\Gamma$ is the bulk Lorentz factor of the outflow along the jet), the angle $\theta_{\rm obs}$ between LOS and the jet axis in the comoving frame is likely to be much larger. Specifically, if $\theta^{\ast}_{\rm obs} = 1/\Gamma$, then $\theta_{\rm obs} = 90^{\circ}$. In our fitting, we choose the LOS in the comoving frame at $\theta_{\rm obs} = 90^{\circ}$. As is mentioned above, the PD variations are hardly linked to the flux variations; therefore, an unpolarized turbulent contribution to the total flux is likely to be present. Therefore, we have fitted the averaged SED for the period, which contains a helical component and a constant turbulent contribution, then fit the polarization signatures based on the derived parameters. Also the PA profile is fitted prior to the PD profile, as the former is less affected by the turbulence. The best fit model and data are displayed in Fig. \ref{fig4}.

We assumed a leptonic origin for the SED, so that the low-energy component is dominated by synchrotron, while the high-energy component consists of both SSC and EC contributions. Due to the Monte-Carlo photon tracing, small numerical errors are present, but the overall fitting is very close to the data. The time-dependent polarization profiles are interpreted as alterations in the magnetic topology initiated by the disturbance. The origin of the disturbance can be either shock or shock-initiated magnetic reconnection. Before the emission region encounters the disturbance, the entire region contributes uniformly. Although the helical magnetic field has comparable poloidal and toroidal contributions, due to the axis-symmetry of the emission region and the LOS orientation, the projected poloidal contribution onto the plane of sky is stronger, resulting in a poloidal-dominating polarized flux. When the disturbance moves in, it will alter the local magnetic field to be toroidal-dominating. In addition, it will strengthen the emission in this region by amplifying the local magnetic field strength (shock) or injecting additional non-thermal electrons (reconnection). Due to the LTTEs, only the near side of the flaring region is observed at first, so that the PD gradually drops as the initial poloidal dominance is balanced by the flaring toroidal contribution, and the PA rotates towards toroidal component. At the middle of the event, the flaring region will extend across the emission region, therefore it will receive equal contributions from both the near and the far sides (Fig. \ref{modelsetup}), as seen from the observer, mimicking the initial axis-symmetry, but dominated by toroidal contribution. Since the length of the cylindrical emission region is longer than its diameter, this flaring region will stay for some time, creating a ``step phase'' in both PD and PA profiles, as proposed in \citet{Zhang2015}. After that, the disturbance will completely leave the emission region, and the flaring region moves to the far side, so that the PD reverts back to its initial state in a time-symmetric pattern. The PA instead completes a $180^{\circ}$ rotation to the initial state (notice the $180^{\circ}$ ambiguity), as the projected toroidal component on the near side is opposite to that on the far side.

  We have further investigated a few issues in the polarization fitting. We notice that although the decreasing part of the PD profile is well reproduced, the rising part around MJD 54047 is a bit off. We remind the readers that we have applied a constant unpolarized turbulence in the fitting, which in reality, may actually decrease with time, consequently PD may increase. The PA, however, is generally unaffected. Moreover, the ``step phase'' is indeed necessary and can put a strong constraint on the ratio of the length and diameter of the emission region. We can see in the data that both PD and PA tend to converge at a stable value during the middle of the event. Most importantly, the slopes of the PA rotation before and after the ``step phase'' set a stringent constraint on the ratio of the flaring toroidal dominance to the initial poloidal dominance. This ratio not only leads to a definite excess of toroidal contribution during the ``step phase'', which is shown in the PD profile, but also require the helical component to undergo an increment in the flux. This is consistent with the flare activity observed at X-rays and $\gamma$-rays. During the PA rotation, the total optical flux is almost stabilized to its highest observed flux and hardly shows any pattern. Therefore, to compromise the flare in the helical component, the turbulent contribution has to decrease, further confirming our previous assumption.

 \section{Discussion} 
 The observations of simultaneous variations in multi-wavelength bands shown in Fig. 1(a)-1(f) suggest that the similar activity is driving the emission over a broad range of spectrum during the brightness phase of \source. The fact that the flare is seen almost simultaneously over a broad spectrum, further suggests that the emission region is likely to be co-spatial. Additionally, it indicates that the region of maximum dissipation should be transparent to the emissions at all energies. The variable VHE detection that followed the activity in optical along with the historically bright state in X-rays and $\gamma$-rays, coinciding with variable PD and rotation of PA, imply that the same electron population is responsible for this major activity. In short, in our model, these observations support a leptonic origin for the high energy emission during the 2015 activity of \source. A careful look at the Fig. \ref{fig1}(a) - \ref{fig1}(f) reveals occurrence of two major sub-flares, in all the bands, super-imposed with small amplitude fluctuation. Any analysis of the time lag between different wavebands is, however, difficult to perform due to limited coverage and their resolution in the present data. The BWB trend and the different slopes in flux-flux correlation graph (Fig. \ref{colvar}) indicate that this flare may have been caused by shock acceleration activities in the jet rather than being triggered by the involvement of geometry dependent effects. The very rapid (hourly) and prominent variations in PD reflect the crucial role played by the magnetic field during this event. The individual segments of PD variations may be due to the fluctuations in the shocked region resulting in changes in the magnetic field in the  compressed region. 
 
  The variations in PA observed during the second optical sub-flare is the most important feature of this state. There is a clear indication of rotation in PA, almost coincident (within a day) with the optical, X-ray and $\gamma$-ray flares. This particular scenario is well observed in some of the blazars namely 3C279, Mrk421, PKS 1510-089 etc. \citep{Abdo2010Nature, Marscher2010, Marscher2014}. Several mechanisms have been proposed to interpret the PA rotations, such as an emission region moving along a curved trajectory \citep{Villata1999} or a bending in the jet \citep{Marscher1991}, or streamlines following the helical magnetic field lines \citep{Marscher2008}, or stochastic activation of individual zones in a turbulent shock (TEMZ, model). The bending jet model involves an asymmetric jet structure and thus requires a lot of freedom in the parameter space. Streamlines along a helical magnetic field imply that the PA rotation should be preferably in the same direction, while the two sequential PA rotations are observed in the opposite directions. Additionally, both models cannot naturally explain the simultaneous flaring activities. The TEMZ model, on the other hand, can hardly produce the systematic, apparently time-symmetric polarization profiles due to its stochastic nature. Nevertheless, our HMFM model applies a simple axisymmetric geometry for the emission region and the disturbance, taking into account all the LTTEs, naturally explaining the simultaneous flaring activities and the apparently time-symmetric polarization profiles. In addition, based on the behavior of the disturbance, an immediately following PA rotation event in the opposite direction is possible. However, since the second PA rotation does not have sufficient data coverage, we are unable to constrain the model parameters. This part is omitted in Fig. \ref{fig4}. 
  
  Our best fit parameters do not allow us to distinguish between the shock scenario and the shock-initiated reconnection scenario. However, we still prefer the reconnection mechanism because our model predicts higher flare amplitude at high energies than in the optical, as is seen here. In the reconnection, the increase of the non-thermal electron density is due to the dissipation of magnetic energy, which weakens the synchrotron flare. The shock scenario, however, enhances both the non-thermal density and the magnetic field strength, leading to a stronger synchrotron flare. The aforementioned flare and PA rotation event is very similar to the other flare + PA rotation events seen in blazars \citep{Abdo2010Nature,Marscher2010,Marscher2014,Zhang2015}. After these events, flux in all bands suffers a large decrease and becomes less active. This indicates that a severe energy dissipation occurs during the flare+PA rotation. The PA rotation implies a strong alteration in the magnetic field, which again provides another piece of evidence for the reconnection.

\section{Summary}
  \source  exhibited a multi-wavelength outburst in January 2015, with two well-resolved sub-flares in gamma-rays and optical. The event was accompanied by rapid PD variations and a systematic PA swing, and followed by a variable VHE detection ($F_{\textgreater 150 GeV}$ = 4-7 $\times$ $10^{-11}$ ph $cm^{-2}$ $s^{-1}$) by the MAGIC group. The total flux and PD variations seem to be uncorrelated, owing to the significant contributions of the unpolarized component to the total emission. However, our study shows the co-spatiality of emission at high and low energies. The observability of quasi-simultaneous emission provides hints about the location of emission region in the jet. The color variations and BWB trend indicate that it may likely be of shock and/or shock-initiated reconnection origin. The HMFM model, adopted to simulate the part of outburst, suggests that the magnetic reconnections more likely played a very important role in this event. The rising part of PA rotation is very well fitted with ``step phase'' profile. The same profile also presents a reasonably good fit for PD variations, except for a slight deviation towards the end. This may be explained by including the contributions of other unpolarized emission components in the optical band. The excess in the ratio of the toroidal component to the poloidal component of the magnetic fields during quiescent and flaring episode indicate the helical component to contribute in the flux increment (Table \ref{tbl2}). In conclusion, our study suggests that this outburst event is more likely to be the similar to 2008 outburst state of \source, and is probably triggered by the shock-initiated magnetic reconnections taking place in the emission region in the jet.


\acknowledgments
The authors are grateful to Prof. P. Smith + team, Arizona University, USA, for making the data from the Steward Observatory spectropolarimetric monitoring project, accessible. This program is supported by Fermi Guest Investigator grants NNX08AW56G, NNX09AU10G, and NNX12AO93G. The authors also acknowledge the HEASARC and Fermi Science Team for the data access from these facilities.  SC and PK  acknowledge the help and support of MIRO local staff during the course of campaign. SC, PK and KPS are thankful to Tata Institute of Fundamental Research, Mumbai for the funding needed for this project. NK and KSB acknowledge the support by Physical Research Laboratory, Unit of Dept. of Space, GOI,  Ahmedabad for partial support. HZ is supported by the LANL/LDRD program and by DoE/Office of Fusion Energy Science through CMSO. MB acknowledges support by the South African Research Chairs Initiative (SARChI) of the Department of Science and Technology and the National Research Foundation of South Africa. The simulations used here, were conducted using LANL's Institutional Computing machines. We thank the referee​ for his positive and encouraging comments.

\clearpage

\begin{figure*}
\centering
\includegraphics[angle=0,width=.85\textwidth, height=.99\textwidth]{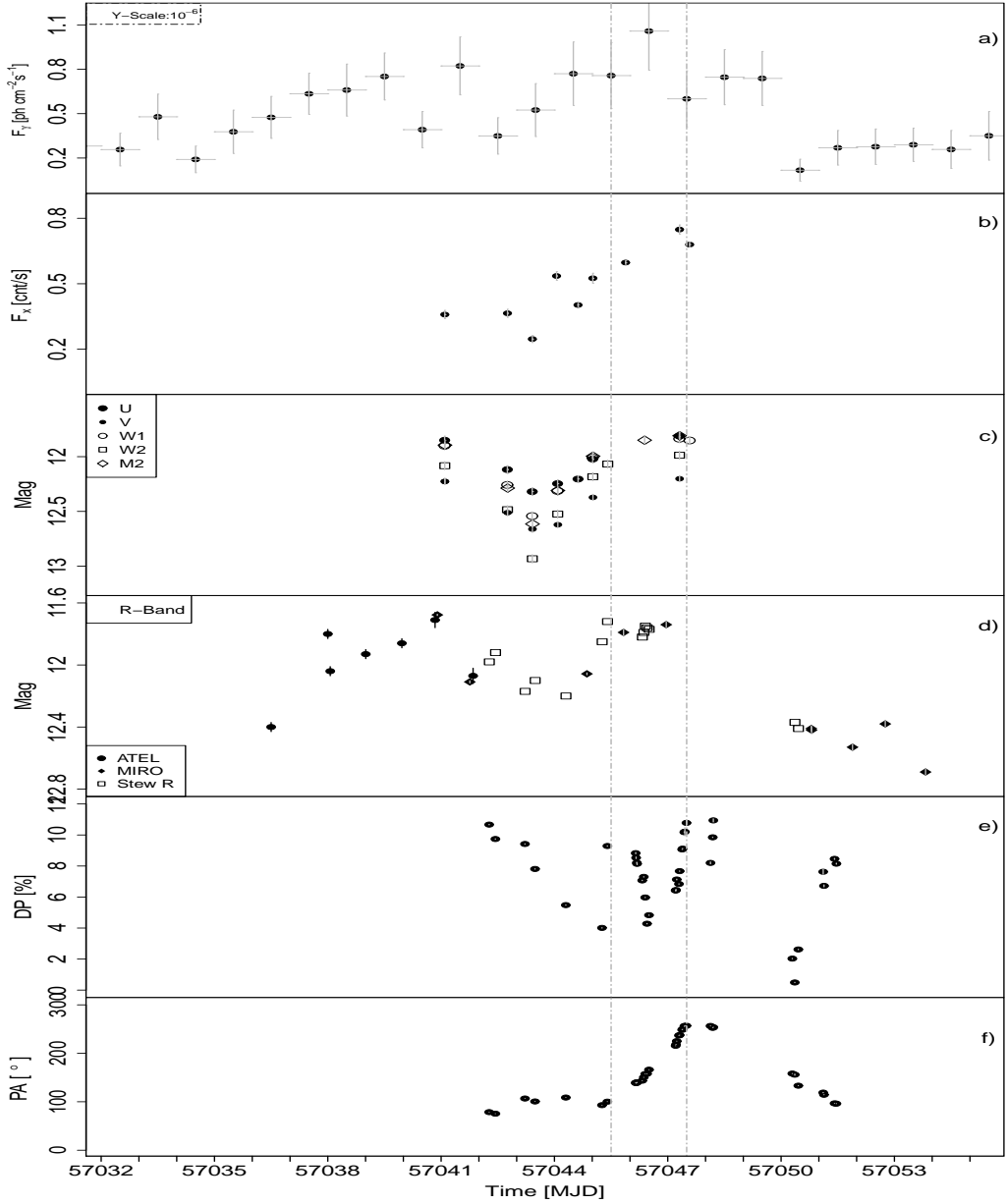}
\caption{Multi-wavelength lightcurve of \source showing the recent outburst activity during 2015, January. Figures 1(a) \& 1(b) respectively represent the Fermi (\textgreater 0.1 GeV) and X-rays (0.3 10.0 keV) light-curves, whereas the Figs. 1(c) \& 1(d) present UV/optical magnitudes from Swift-UVOT and MIRO, respectively. Last two panels (1(e) \& 1(f)) are PD and PA variations. The `Stew R' notation in Fig. 1(d) stands for R band data from Steward Observatory.. \label{fig1}}
\end{figure*}

\begin{figure*}
\includegraphics[angle=0,scale=.48]{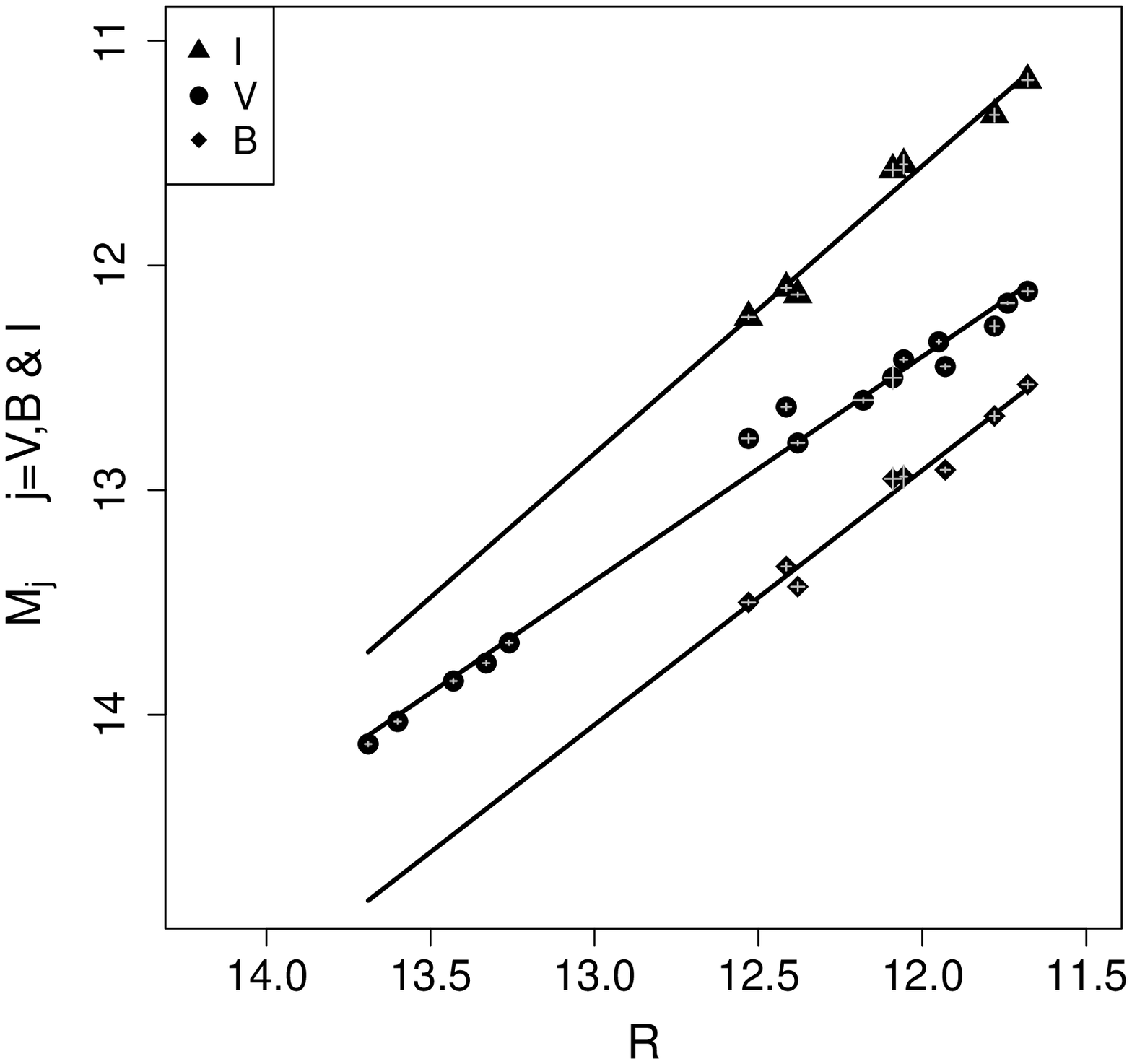}
\includegraphics[scale=0.48]{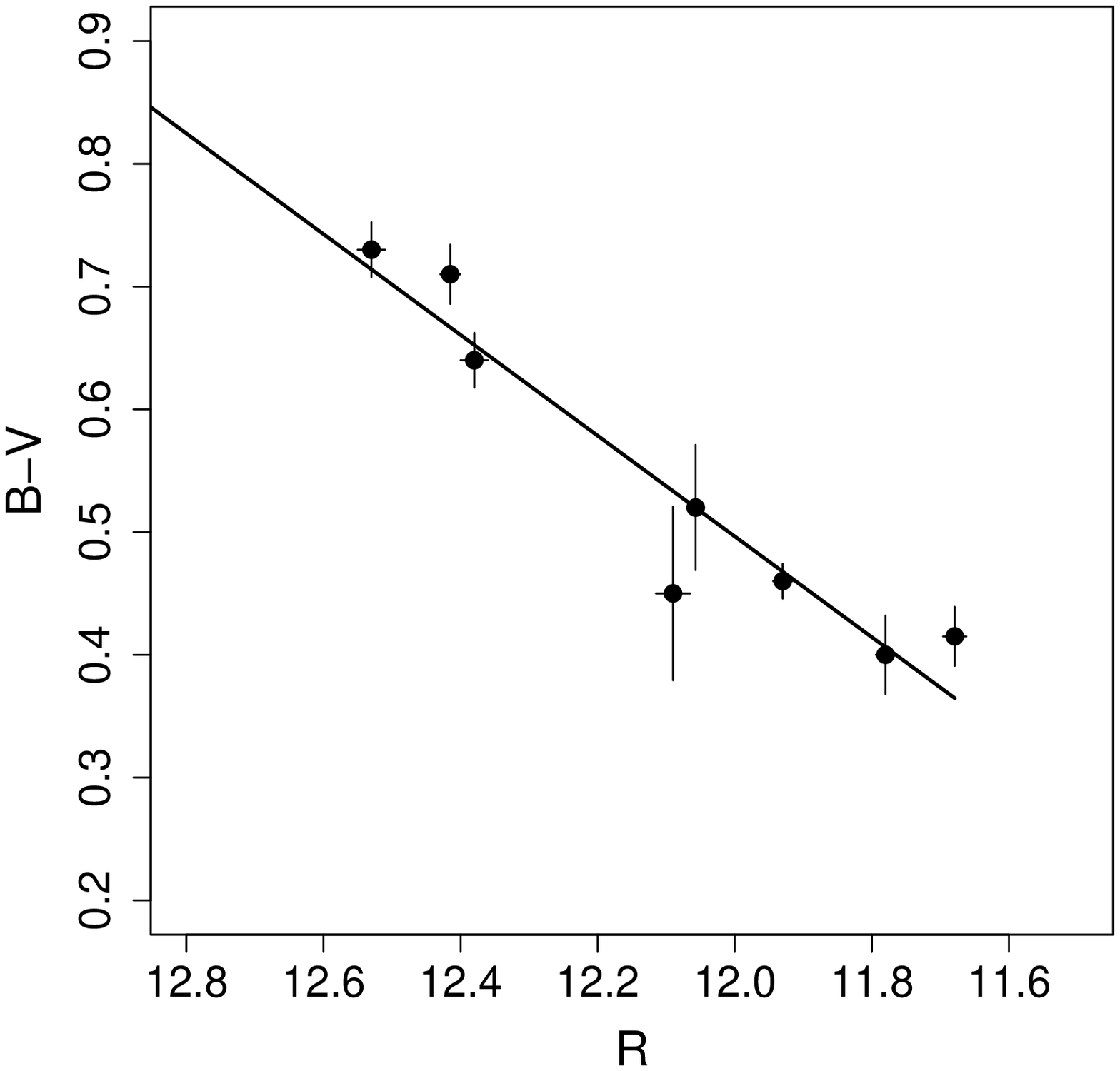}
\caption{{\it Left:} R band and I,V, B optical bands correlations curves for fluxes in different optical bands. The X-axis  shows the R band magnitude whereas the Y-axis represents magnitudes in other optical bands namely, I, B, \& V. {\it Right:} The flux-color plot for optical observations. The X \& Y axes are the R-band magnitude and (B-V) color, respectively.  \label{colvar}}
\end{figure*}   

\begin{figure*}
\includegraphics[angle=0,scale=.68]{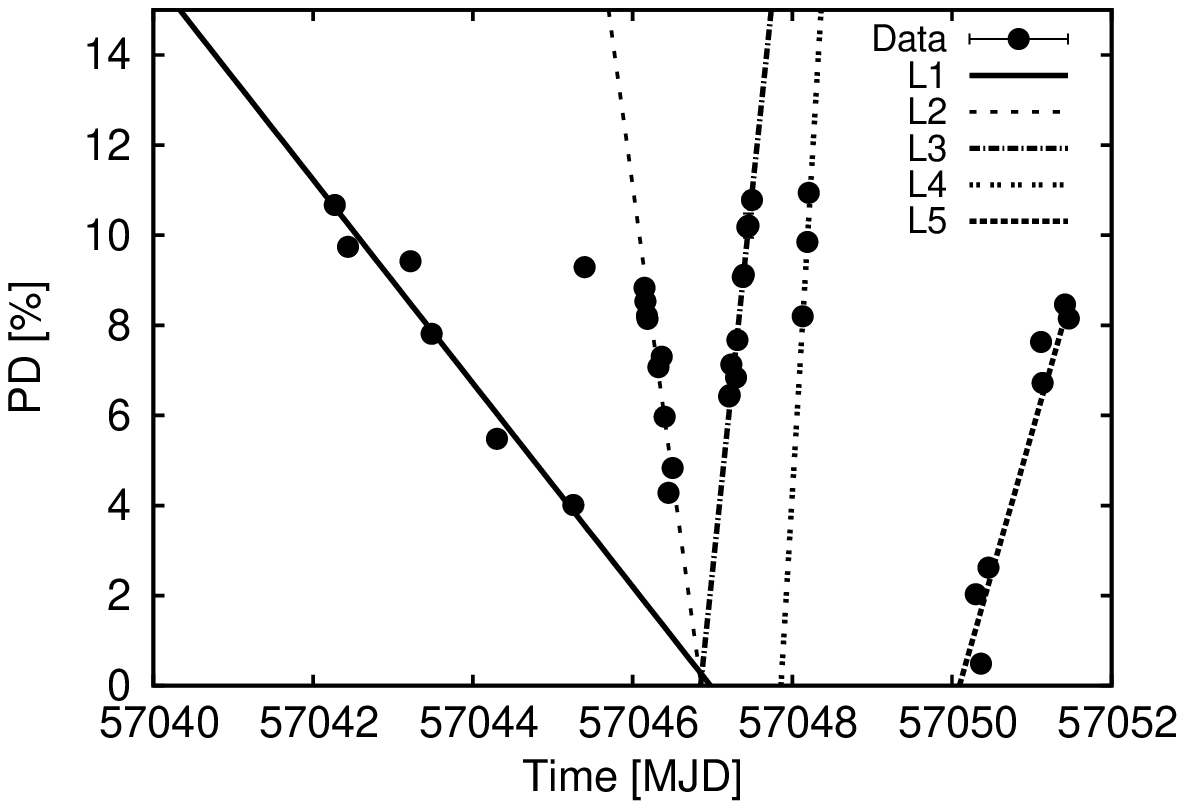}
\includegraphics[angle=0,scale=.68]{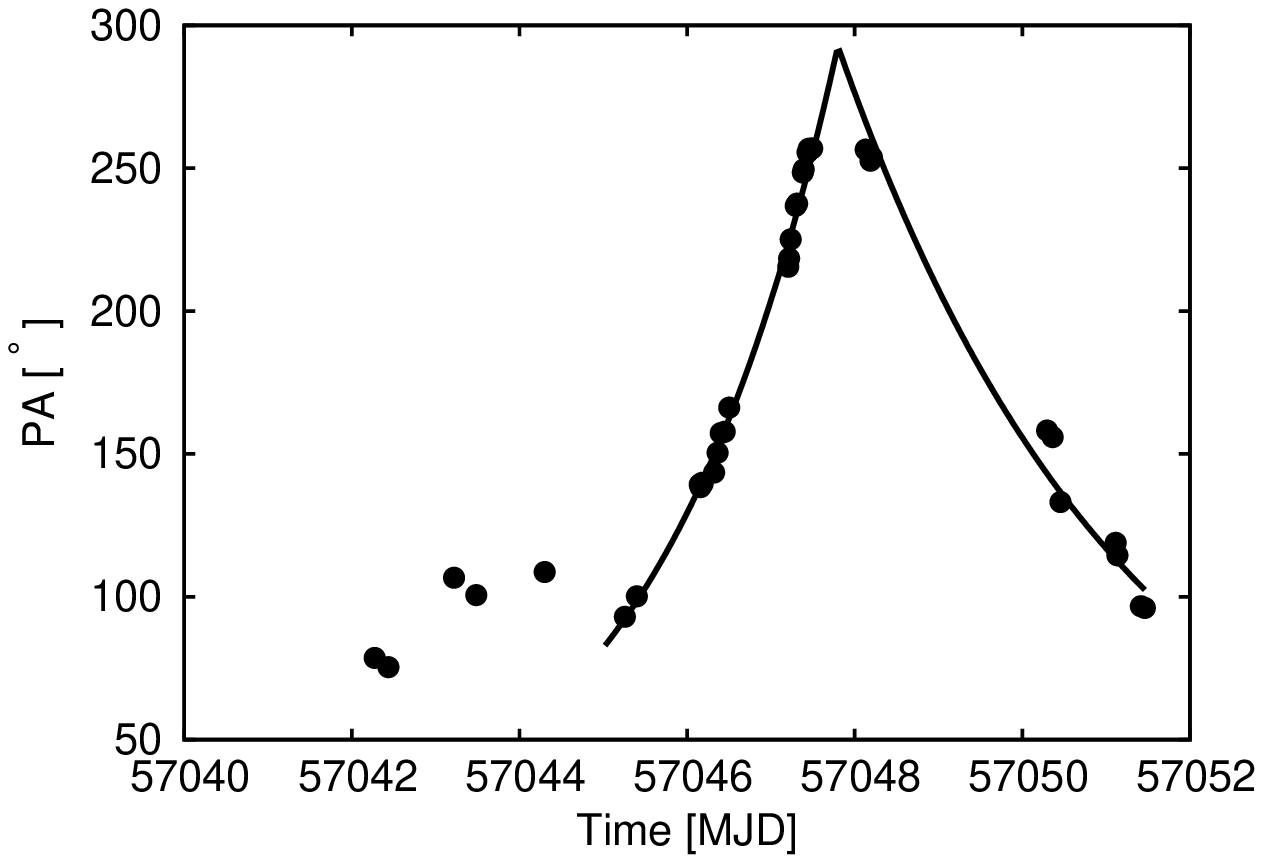}
\caption{{\it Left:} The PD variation with individual segments fitted by straight lines (L1 to L5). The fitting is performed using least square fitting algorithm. {\it Right:} The PA variations fitted with an exponential rising and falling profile given by [$f(x) \sim A e^{-\alpha (x-x0)}$].\label{pollinfit}}
\end{figure*}

\begin{figure*}
\centering
\includegraphics[width=0.9\textwidth]{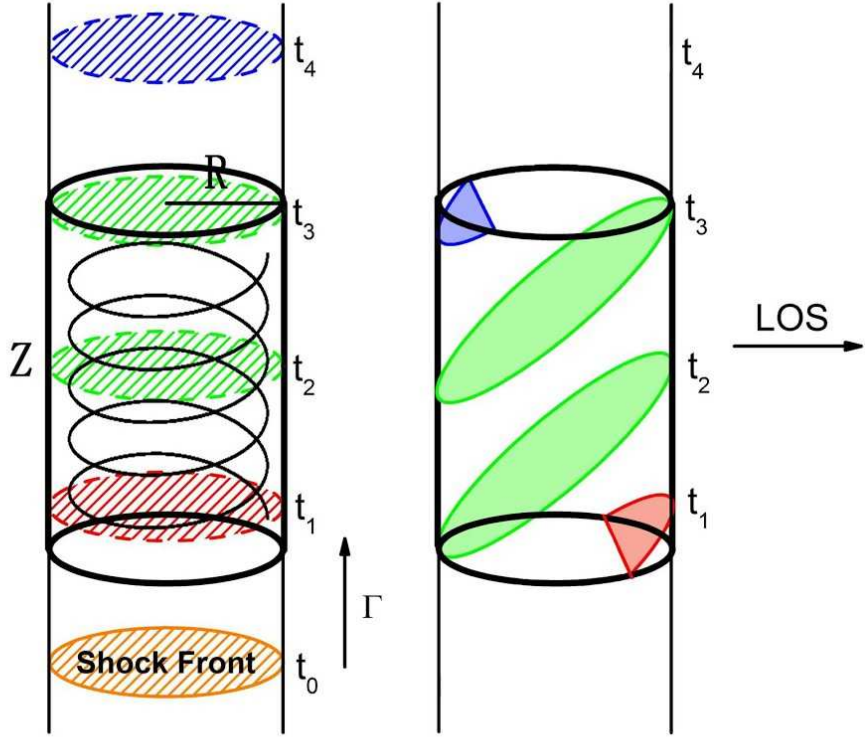}
\caption{Left: Sketch of the interaction between the emission region and the disturbance in the comoving frame of the emission region, at different epochs. The emission region is pervaded by a helical magnetic field and a turbulent component (only the helical component is sketched). The disturbance is stationary in the observer's frame, but in the comoving frame of the emission region, the disturbance is then moving up with Lorentz factor $\Gamma$. The orange, red, green and blue regions refer to the locations of the disturbance before the flare ($t_0$), the rising phase ($t_1$), peak ($t_2$ and $t_3$) and declining phase ($t_4$), respectively. Right: The red, green and blue shapes indicate the shape and location of the flaring region, corresponding to the disturbance at $t_1\sim t_3$, respectively, observed simultaneously, taking into account the LTTEs. Since $Z>2R$, the peak state will stay for few hours or few days, depending upon the Z/R ratio \citep{Zhang2015}.
\label{modelsetup}}
\end{figure*}

\begin{figure*}
\includegraphics[angle=0,scale=.68]{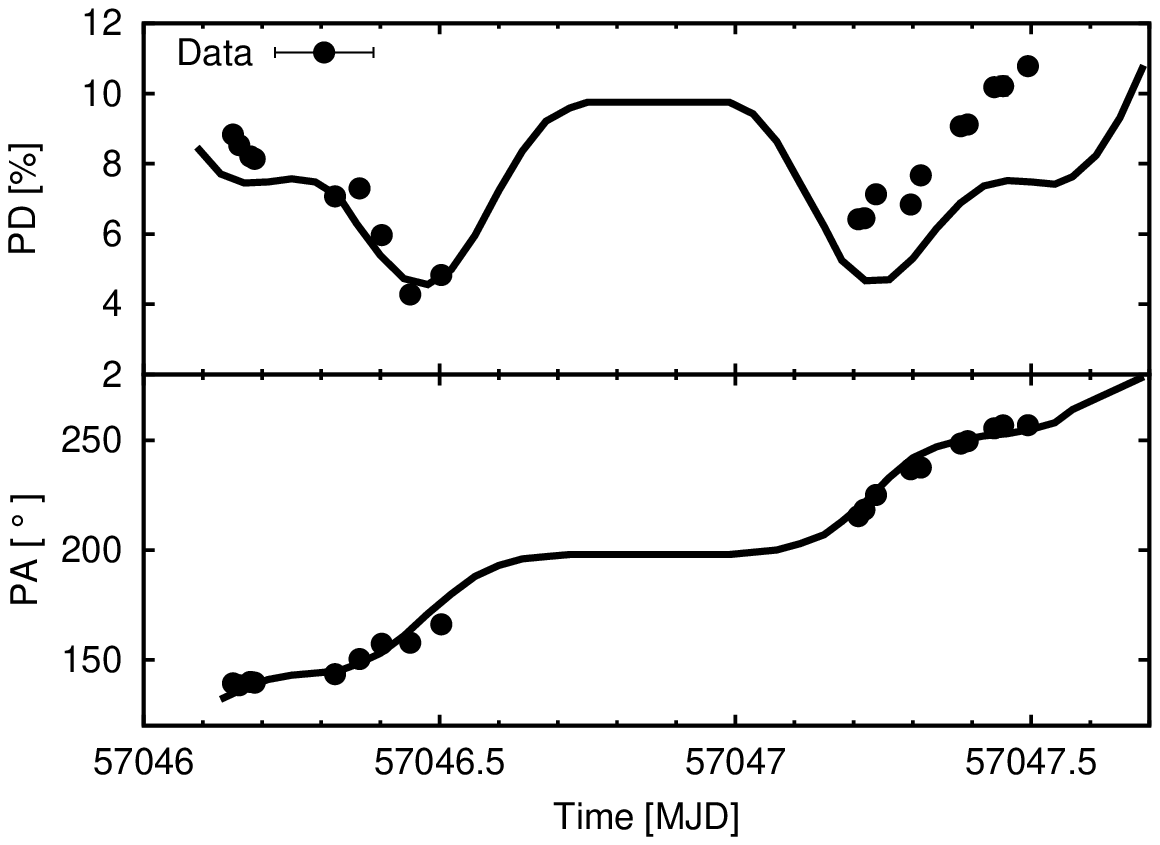}
\includegraphics[angle=0,scale=.68]{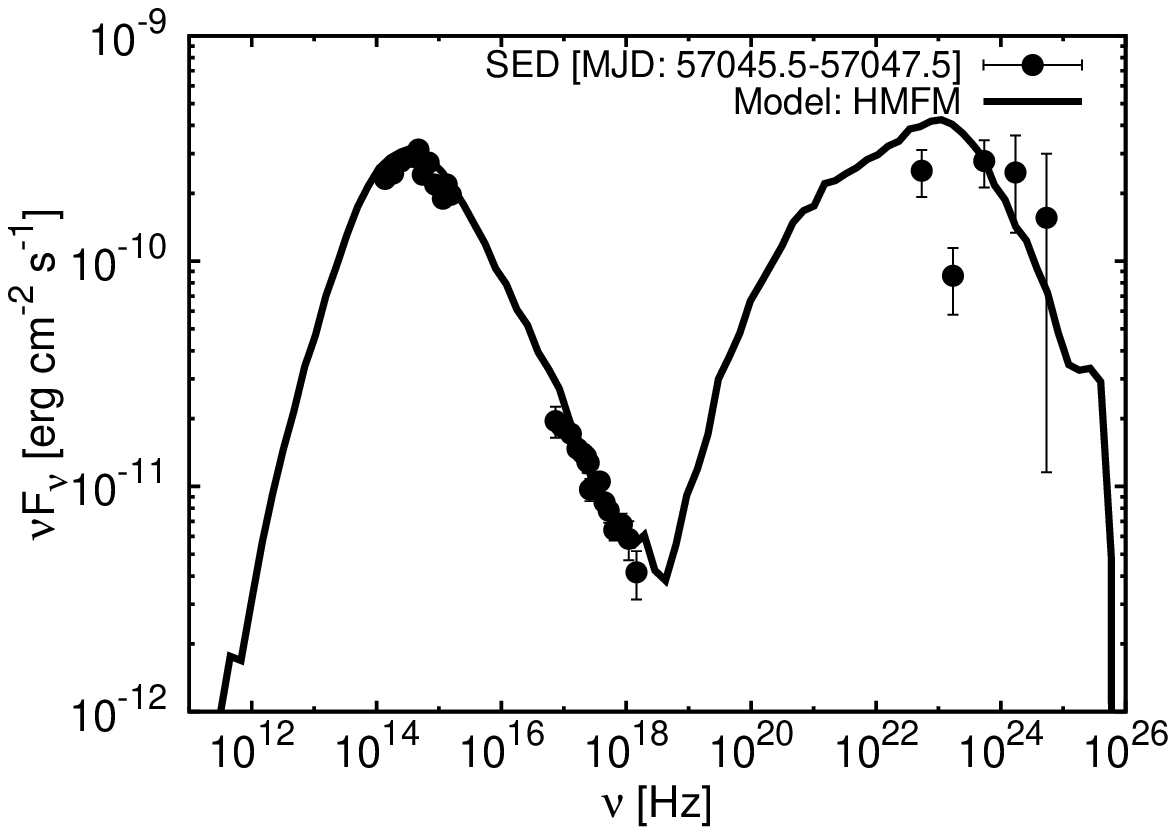}
\caption{{\it Left:} The model reproduced PD and PA, generated for the duration of MJD 57045.5 to 57047.5. The PD variations overplotted with the modeled curve (thick black line). The bottom panel represents the data and model for PA swing. {\it Right:} The broadband SED of \source for aforementioned duration. The multi-wavelength light-curves and respective models are not shown here as the flux remains almost constant (though highest) for the duration under consideration.\label{fig4}}
\end{figure*}

\begin{table*}
\caption{Swift XRT Spectral Fitting Parameters. \label{tbl1}}
\begin{tabular}{lclcllc}
\hline
\hline
{\it ObsID} & {\it Exposure} & Time & {\it Sp. Index ($\Gamma$)} & {\it $log_{10}$($F_{0.3-10.0 keV}$)} & $\chi^2$ & dof \\
&(s)&(MJD)&&(erg $cm^{-2}$ $s^{-1}$) & & \\
\hline
00035009145  &  1043  &  57023.21  & $ 2.3 \pm 0.4  $& $-11.4 \pm 0.10 $ &  4.6  &  3   \\
00035009146  &  821   &  57029.01  & $ 1.9 \pm 0.4  $& $-11.3 \pm 0.13 $ &  1.6  &  2   \\
00035009147  &  976   &  57041.09  & $ 2.7 \pm 0.3  $& $-11.0 \pm 0.06 $ &  5.0  &  11  \\
00035009148  &  1091  &  57042.75  & $ 2.4 \pm 0.2  $& $-10.9 \pm 0.06 $ &  14.1	&  13  \\
00035009149  &  961   &  57043.41  & $ 1.8 \pm 0.6  $& $-10.7 \pm 0.20 $ &  1.7  &  3   \\
00035009152  &  1348  &  57044.02  & $ 2.3 \pm 0.2  $& $-10.7 \pm 0.05 $ &  23.1  &  19  \\
00035009153  &  6895  &  57044.29  & $ 2.3 \pm 0.1  $& $-10.8 \pm 0.03 $ &  38.4  &  50  \\
00035009154  &  978   &  57045.01  & $ 2.5 \pm 0.3  $& $-10.6 \pm 0.06 $ &  6.6  &  9   \\
00035009156  &  9574  &  57045.14  & $ 2.5 \pm 0.1  $& $-10.7 \pm 0.02 $ &  92.2  &  89  \\
00035009157  &  1688  &  57047.14  & $ 2.6 \pm 0.1  $& $-10.6 \pm 0.03 $ &  17.2	&  26  \\
00035009158  &  6557  &  57047.22  & $ 2.6 \pm 0.1  $& $-10.6 \pm 0.02 $ &  79.8  &  68  \\
\hline

\end{tabular}

\end{table*}

\begin{table*}
\caption{The best estimates of the model parameters. \label{tbl2}}
\begin{tabular}{|l|c|}
\hline
\hline
Paramters & Value (CGS) \\
\hline	
Bulk Lorentz factor & 	20 \\
Length of the emission region Z (cm)	 & 6.06 $\times 10^{16}$ \\
Radius of the emission region R (cm)	 & 2.25 $\times 10^{16}$ \\
Length of the disturbance L (cm)	     & 6.06 $\times 10^{15}$ \\
Radius of the disturbance A (cm)      &	2.25 $\times 10^{16}$ \\
Orientation of LOS (deg)	             & 90 \\
Electron acceleration time-scale (Z/c) &	5.50 $\times 10^{-3}$ \\
Electron escape time-scale (Z/c)	  &  6.00 $\times 10^{-4}$ \\
Electron density ($cm^{-3}$)   & 	21.7 \\
Helical magnetic field strength (G) & 	0.5 \\
Helical pitch angle (deg)	&  47 \\
Helical pitch angle during flare (deg)	& 75.5 \\
\hline
\end{tabular}
\end{table*}

%
%

\clearpage


\newpage
\clearpage
\begin{thebibliography}{}

\bibitem[Abdo et al.(2009)]{Abdo2009} Abdo, A.~A., Ackermann, 
M., Ajello, M., et al.\ 2009, \apj, 700, 597 

\bibitem[Abdo et al.(2010)]{Abdo2010Nature} Abdo, A.~A., Ackermann, 
M., Ajello, M., et al.\ 2010, \nat, 463, 919 

\bibitem[Agudo et al.(2011)]{Agudo2011} Agudo, I., Marscher, 
A.~P., Jorstad, S.~G., et al.\ 2011, \apjl, 735, L10 

\bibitem[Aharonian et 
al.(2004)]{Aharonian2004} Aharonian, F., Akhperjanian, A., Beilicke, M., et al.\ 2004, \aap, 421, 529 

\bibitem[Albert et al.(2006)]{Albert2006} Albert, J., Aliu, E., 
Anderhub, H., et al.\ 2006, \apjl, 648, L105 

\bibitem[Albert et al.(2007)]{Albert2007a} Albert, J., Aliu, E., 
Anderhub, H., et al.\ 2007, \apjl, 666, L17 

\bibitem[Anderhub et al.(2009)]{Anderhub2009} Anderhub, H., 
Antonelli, L.~A., Antoranz, P., et al.\ 2009, \apjl, 704, L129 

\bibitem[Arkharov et al.(2015)]{Arkharov6942Atel} Arkharov, A.~A., 
Borman, G.~A., Di Paola, A., Larionov, V.~M., 
\& Morozova, D.~A.\ 2015, The Astronomer's Telegram, 6942, 1 

\bibitem[Atwood et al.(2009)]{Atwood2009} Atwood, W.~B., Abdo, 
A.~A., Ackermann, M., et al.\ 2009, \apj, 697, 1071 

\bibitem[Bachev et al.(2015)]{Bachev6944Atel} Bachev, R., Spassov, B., 
\& Boeva, S.\ 2015, The Astronomer's Telegram, 6944, 1 

\bibitem[Bachev 
\& Strigachev(2015)]{Bachev6957Atel} Bachev, R., \& Strigachev, A.\ 2015, The Astronomer's Telegram, 6957, 1 

\bibitem[Bessell(1979)]{Bessell1979PASP} Bessell, M.~S.\ 1979, \pasp, 
91, 589 

\bibitem[B{\l}a{\.z}ejowski et al.(2000)]{Blazejowski2000} 
B{\l}a{\.z}ejowski, M., Sikora, M., Moderski, R., \& Madejski, G.~M.\ 2000, \apj, 545, 107 

\bibitem[Bloom 
\& Marscher(1996)]{BloomMarscher1996} Bloom, S.~D., \& Marscher, A.~P.\ 1996, \apj, 461, 657 

\bibitem[Burrows et al.(2005)]{Burrows2005} Burrows, D.~N., Hill, 
J.~E., Nousek, J.~A., et al.\ 2005, \ssr, 120, 165

\bibitem[Cardelli et al.(1989)]{Cardelli1989} Cardelli, J.~A., 
Clayton, G.~C., \& Mathis, J.~S.\ 1989, \apj, 345, 245 

\bibitem[Carrasco et al.(2015)]{Carrasco7268Atel} Carrasco, L., Porras, 
A., Recillas, E., et al.\ 2015, The Astronomer's Telegram, 7268, 1 

\bibitem[Carrasco et al.(2015)]{Carrasco7026ATel.7026....1C} Carrasco, L., Porras, 
A., Recillas, E., et al.\ 2015, The Astronomer's Telegram, 7026, 1 

\bibitem[Carrasco et al.(2015)]{Carrasco6902Atel} Carrasco, L., Porras, 
A., Recillas, E., et al.\ 2015, The Astronomer's Telegram, 6902, 1 

\bibitem[Chandra et al.(2011)]{Chandra2011} Chandra, S., Baliyan, 
K.~S., Ganesh, S., \& Joshi, U.~C.\ 2011, \apj, 731, 118 

\bibitem[Chandra et al.(2012)]{Chandra2012} Chandra, S., Baliyan, 
K.~S., Ganesh, S., \& Joshi, U.~C.\ 2012, \apj, 746, 92 

\bibitem[Chandra et al.(2014)]{Chandra2014} Chandra, S., Baliyan, 
K.~S., Ganesh, S., \& Foschini, L.\ 2014, \apj, 791, 85 

\bibitem[Chandra et al.(2015)]{Chandra6962Atel} Chandra, S., Kushwah, 
P., Ganesh, S., Kaur, N., 
\& Baliyan, K.\ 2015, The Astronomer's Telegram, 6962, 1 

\bibitem[Chen et 
al.(2008)]{Chen2008} Chen, A.~W., D'Ammando, F., Villata, M., et al.\ 2008, \aap, 489, L37 

\bibitem[Chen et al.(2012)]{Chen2012} Chen, X., Fossati, G., 
B{\"o}ttcher, M., \& Liang, E.\ 2012, \mnras, 424, 789 

\bibitem[Dermer et 
al.(1992)]{Dermer1992} Dermer, C.~D., Schlickeiser, R., \& Mastichiadis, A.\ 1992, \aap, 256, L27 

\bibitem[Ferrero et 
al.(2006)]{Ferrero2006} Ferrero, E., Wagner, S.~J., Emmanoulopoulos, D., \& Ostorero, L.\ 2006, \aap, 457, 133 

\bibitem[Foschini et 
al.(2006)]{Foschini2006} Foschini, L., Tagliaferri, G., Pian, E., et al.\ 2006, \aap, 455, 871 

\bibitem[Ganesh et al.(2009)]{Ganesh2009} Ganesh, S., Joshi, 
U.~C., Baliyan, K.~S., et al.\ 2009, arXiv:0912.0076 

\bibitem[Ganesh et al.(2015)]{Ganesh2015} Ganesh, S., Baliyan, 
K.~S., Mishra, A., et al.\ 2015, Under Preperation.

\bibitem[Gehrels et al.(2005)]{Gehrels2005} Gehrels, N., 
Chincarini, G., Giommi, P., et al.\ 2005, \apj, 621, 558 

\bibitem[Gehrels et al.(2004)]{Gehrels2004} Gehrels, N., 
Chincarini, G., Giommi, P., et al.\ 2004, \apj, 611, 1005 

\bibitem[Giommi et 
al.(1999)]{Giommi1999} Giommi, P., Massaro, E., Chiappetti, L., et al.\ 1999, \aap, 351, 59 

\bibitem[Giommi et al.(2008)]{Giommi2008} Giommi, P., Perri, M., 
Verrecchia, F., et al.\ 2008, The Astronomer's Telegram, 1495, 1 

\bibitem[Ghisellini et 
al.(1985)]{Ghisellini1985} Ghisellini, G., Maraschi, L., \& Treves, A.\ 1985, \aap, 146, 204 

\bibitem[Ghisellini et al.(2009)]{Ghisellini2009} Ghisellini, G., 
Tavecchio, F., \& Ghirlanda, G.\ 2009, \mnras, 399, 2041 

\bibitem[Hartman et al.(1999)]{Hartman1999} Hartman, R.~C., 
Bertsch, D.~L., Bloom, S.~D., et al.\ 1999, \apjs, 123, 79 

\bibitem[Kalberla et 
al.(2005)]{Kalberla2005} Kalberla, P.~M.~W., Burton, W.~B., Hartmann, D., et al.\ 2005, \aap, 440, 775 

\bibitem[Kalberla et 
al.(2010)]{Kalberla2010} Kalberla, P.~M.~W., McClure-Griffiths, N.~M., Pisano, D.~J., et al.\ 2010, \aap, 521, A17 

\bibitem[Kushwaha et al.(2014)]{Kushwaha2014} Kushwaha, P., Singh, 
K.~P., \& Sahayanathan, S.\ 2014, \apj, 796, 61 

\bibitem[Larionov et al.(2008)]{Larionov2008} Larionov, V., 
Konstantinova, T., Kopatskaya, E., et al.\ 2008, The Astronomer's Telegram, 
1502, 1 

\bibitem[Marscher et al.(1991)]{Marscher1991} Marscher, A.~P., Zhang, Y.~F., Shaffer, D.~B., Aller, H.~D., \& Aller, M.~F.\ 1991, \apj, 371, 491 

\bibitem[Marscher et al.(2008)]{Marscher2008} Marscher, A.~P., 
Jorstad, S.~G., D'Arcangelo, F.~D., et al.\ 2008, \nat, 452, 966 

\bibitem[Marscher(2014)]{Marscher2014} Marscher, A.~P.\ 2014, \apj, 
780, 87 

\bibitem[Marscher et al.(2010)]{Marscher2010} Marscher, A.~P., 
Jorstad, S.~G., Larionov, V.~M., et al.\ 2010, \apjl, 710, L126 

\bibitem[McClure-Griffiths et al.(2009)]{McClure2009} 
McClure-Griffiths, N.~M., Pisano, D.~J., Calabretta, M.~R., et al.\ 2009, 
\apjs, 181, 398 

\bibitem[Mirzoyan(2015)]{Mirzoyan6999Atel} Mirzoyan, R.\ 2015, The 
Astronomer's Telegram, 6999, 1 

\bibitem[Nalewajko et al.(2014)]{Nalewajko2014} Nalewajko, K., 
Sikora, M., \& Begelman, M.~C.\ 2014, \apjl, 796, L5 

\bibitem[Nandikotkur et al.(2007)]{Nandikotkur2007} Nandikotkur, G., 
Jahoda, K.~M., Hartman, R.~C., et al.\ 2007, \apj, 657, 706 

\bibitem[Nilsson et 
al.(2008)]{Nilsson2008} Nilsson, K., Pursimo, T., Sillanp{\"a}{\"a}, A., Takalo, L.~O., \& Lindfors, E.\ 2008, \aap, 487, L29 

\bibitem[Nolan et al.(2012)]{Nolan2012} Nolan, P.~L., Abdo, 
A.~A., Ackermann, M., et al.\ 2012, \apjs, 199, 31 

\bibitem[Orienti et al.(2013)]{Orienti2013} Orienti, M., Koyama, 
S., D'Ammando, F., et al.\ 2013, \mnras, 428, 2418 

\bibitem[Qian et al.(2002)]{Qian2002} Qian, B., Tao, J., 
\& Fan, J.\ 2002, \aj, 123, 678 

\bibitem[Raiteri et 
al.(2003)]{Raiteri2003} Raiteri, C.~M., Villata, M., Tosti, G., et al.\ 2003, \aap, 402, 151 

\bibitem[Roming et al.(2005)]{Roming2005} Roming, P.~W.~A., 
Kennedy, T.~E., Mason, K.~O., et al.\ 2005, \ssr, 120, 95 

\bibitem[Schlafly 
\& Finkbeiner(2011)]{Schlafly2011} Schlafly, E.~F., \& Finkbeiner, D.~P.\ 2011, \apj, 737, 103 

\bibitem[Sikora et al.(1994)]{Sikora1994} Sikora, M., Begelman, 
M.~C., \& Rees, M.~J.\ 1994, \apj, 421, 153 


\bibitem[Sikora et al.(2009)]{Sikora2009} Sikora, M., Stawarz, 
{\L}., Moderski, R., Nalewajko, K., \& Madejski, G.~M.\ 2009, \apj, 704, 38 

\bibitem[Spiridonova et al.(2015)]{Spiridonova6953Atel} Spiridonova, O.~I., 
Vlasyuk, V.~V., Moskvitin, A.~S., 
\& Bychkova, V.~S.\ 2015, The Astronomer's Telegram, 6953, 1 

\bibitem[Sokolov et al.(2004)]{Sokolov2004} Sokolov, A., Marscher, 
A.~P., \& McHardy, I.~M.\ 2004, \apj, 613, 725 

\bibitem[Smith et al.(2009)]{Smith2009} Smith, P.~S., Montiel, 
E., Rightley, S., et al.\ 2009, arXiv:0912.3621 

\bibitem[Urry 
\& Padovani(2000)]{UrryPadovani2000} Urry, M., \& Padovani, P.\ 2000, \pasp, 112, 1516 

\bibitem[Urry 
\& Padovani(1995)]{UrryPadovani1995} Urry, C.~M., \& Padovani, P.\ 1995, \pasp, 107, 803 

\bibitem[Urry 
\& Mushotzky(1982)]{UrryMushotzky1982} Urry, C.~M., \& Mushotzky, R.~F.\ 1982, \apj, 253, 38 

\bibitem[Villata 
\& Raiteri(1999)]{Villata1999} Villata, M., \& Raiteri, C.~M.\ 1999, BL Lac Phenomenon, 159, 489 

\bibitem[Villata et 
al.(2004)]{Villata2004} Villata, M., Raiteri, C.~M., Kurtanidze, O.~M., et al.\ 2004, \aap, 421, 103 

\bibitem[Wagner et al.(1996)]{WagnerWitzel1996} Wagner, S.~J., Witzel, 
A., Heidt, J., et al.\ 1996, \aj, 111, 2187 

\bibitem[Wagner 
\& Witzel(1995)]{Wagner1995} Wagner, S.~J., \& Witzel, A.\ 1995, \araa, 33, 163 

\bibitem[Zhang et al.(2014)]{Haocheng2014} Zhang, H., Chen, X., {\ 
B\"o}ttcher, M., 2014, \apj, 789, 66 

\bibitem[Zhang et al.(2015)]{Zhang2015} Zhang, H., Chen, X., 
B{\"o}ttcher, M., Guo, F., \& Li, H.\ 2015, \apj, 804, 58 

  
\end{thebibliography}
\end{document}